\documentclass[10pt,conference]{IEEEtran}

\usepackage{amsmath}
\usepackage[english]{babel}
\usepackage{blindtext}
\usepackage{booktabs} 
\usepackage{multirow}
\usepackage{caption}
\usepackage[FIGBOTCAP]{subfigure}
\usepackage{adjustbox}
\usepackage{enumitem}
\usepackage{colortbl}
\usepackage[normalem]{ulem}
\usepackage{soul}
\usepackage{balance}

\newcommand{\ie}{{\em i.e., }}
\newcommand{\eg}{{\em e.g., }}
\newcommand{\myverb}{\fontsize{9}{48}\usefont{OT1}{lmtt}{b}{n}\noindent }
\usepackage{tcolorbox} 
\newcommand{\ciao}[1]{{\setlength\fboxrule{0pt}\fbox{\tcbox[colframe=black,colback=white,shrink tight,boxrule=0.5pt,extrude by=1mm]{\small #1}}}}

\def\BibTeX{{\rm B\kern-.05em{\sc i\kern-.025em b}\kern-.08emT\kern-.1667em\lower.7ex\hbox{E}\kern-.125emX}}

\begin{document}
	
\title{Realizing Open and Decentralized Marketplace for Exchanging Data of Expected IoT Behaviors\\(Full Version)}

\author{
	\IEEEauthorblockN{Song Guo, Minzhao Lyu and Hassan Habibi Gharakheili }
	\IEEEauthorblockA{
		School of Electrical Engineering and Telecommunications, UNSW Sydney, Australia\\
		Emails: \textit{song.guo@unswalumni.com, \{minzhao.lyu@, h.habibi@\}unsw.edu.au}}
}

\maketitle
\begin{abstract}

With rising concerns about the security of IoT devices, network operators need better ways to handle potential risks. Luckily, IoT devices show consistent patterns in how they communicate. But despite previous efforts, it remains unclear how knowledge of these patterns can be made available.
As data marketplaces become popular in different domains, this paper\footnote{This manuscript is the full version of our paper \cite{PubVersion} accepted to the IEEE/IFIP NOMS 2024 conference.} proposes creating a special marketplace focused on IoT cybersecurity. The goal is to openly share knowledge about IoT devices' behavior, using structured data formats like Manufacturer Usage Description (MUD) files. To make this work, we employ technologies like blockchain and smart contracts to build a practical and secure foundation for sharing and accessing important information about how IoT devices should behave on the network.
Our contributions are two-fold. 
(1) We identify the essential features of an effective marketplace for sharing data related to the expected behaviors of IoT devices. We develop a smart contract on the Ethereum blockchain with five concrete functions; and,
(2) We implement a prototype of our marketplace in a private chain environment---our codes are publicly released. We demonstrate how effectively our marketplace functions through experiments involving MUD files from consumer IoT devices. Our marketplace enables suppliers and consumers to share MUD data on the Ethereum blockchain for under a hundred dollars, promoting accessibility and participation.
\end{abstract}
\vspace{2mm}

\section{Introduction} \label{intro}
Networks rich in connected IoT devices, such as cameras, sensors, lightbulbs, and printers, remain attractive to cyber attackers due to a lack of sufficient built-in security measures \cite{SGreengardCA2019,ESchillerCSR2022}, primarily driven by factors such as low-profit margins, inexperience/unincentivized manufacturers, low-cost design choices, or limited computing resources. This, plus the heterogeneity of hardware and software components across various IoT device types, make managing cyber risks at scale even more challenging.

While certain (large and mature) manufacturers continue to invest in implementing device-specific security at various layers, a network-level approach, given the identifiable patterns and repeatable nature of IoT behaviors \cite{ASivanathanTMC2019}, is perceived as complementary and more systematic in protecting the entire range of connected devices. Among several proposals, the IETF standard for Manufacturer Usage Description (MUD) \cite{rfcMUD} has received attention from industry and academia due to its enforceable \cite{18iotsnpMUDids} and relatively simple data structure \cite{MUDcheck2022} to specify the expected behaviors of IoT devices. The MUD standard: (a) specifies a ``data model'' (JSON-formatted) to describe a set of access control entries representing allowed communications of an IoT device on the network, and (b) defines in-band mechanisms (\eg DHCP, LLDP) by which an IoT device can ``signal'' to the network supplying (a MUD URL to) the data model that indicates what access and protection the device needs.

Though the MUD data can reduce attack surfaces \cite{AHamzaSOSR2019} and help manage the health of connected IoT devices \cite{APashamokhtariWoWMoM2022}, not many manufacturers are motivated to supply it. This is mainly because signaling (the MUD URL) to the network requires modifications to the device firmware, but currently, no network gateway automatically consumes those URLs \cite{rfc9238}. In addition, many manufacturers use open-source and/or third-party components and, therefore, may lack full knowledge \cite{NISTIR8349} of the expected behaviors for IoT devices they produce---unable (less incentivized) to publish the corresponding MUD data. Also, it is important to note that MUD data (even published in the first instance) is subject to change at slower time scales (say, months or years) due to device variations or firmware upgrades. Therefore, data needs updates too.  

IoT users (network operators) can obtain knowledge and generate data \cite{MUDcheck2022} on device behaviors by measuring network activities.
This paper advocates an open and decentralized platform over which various entities, such as network operators, cybersecurity researchers, compliance-check laboratories, as well as device manufacturers, can dynamically exchange the data of expected device behaviors, helping to secure IoT networks at scale. 
We build upon prior relevant work and focus on developing an API-driven marketplace for automatically requesting, competitively supplying, and explicitly rating MUD files. 

Our specific contributions are twofold.
\textbf{First}, we identify key features for the open and immutable sharing of IoT expected behavior data, promoting systematic collaboration within the cybersecurity community. This effort involves device manufacturers, consumers, security researchers, and service providers, aiming to replace the current non-transparent practices on centralized data marketplaces.
Based on the identified features, we conceptualize a decentralized data marketplace that operates on the Ethereum blockchain. Our solution includes a specialized smart contract with five key functions: request, offer, select, share, and rate. Additionally, we incorporate off-chain storage for MUD files, allowing every marketplace node to contribute and access immutable MUD data.
\textbf{Second}, we implement a prototype of our data marketplace on a private Ethereum instance and evaluate its efficacy by experimenting with publicly available MUD files of several consumer IoT devices. We contribute our code (smart contract implementation and experimental scenarios) as open source \cite{shareMUDrepo2023} for the research community to use.

The rest of this paper is organized as follows. We discuss relevant prior work in \S\ref{sec:related}. Our conceptual design of the data marketplace for sharing IoT MUD data is presented in \S\ref{sec:tool}. We implement and evaluate our prototype in \S\ref{sec:eval}. We highlight the limitations of our work and outline potential future directions in \S\ref{sec:discuss}. The paper is concluded in \S\ref{sec:conclude}.

\section{Related Work}\label{sec:related}
Let us begin by examining prior relevant work in the field, particularly in the development of data marketplaces for IoT devices and their use of MUD data for IoT network security. Additionally, we assess current commercial data marketplaces, highlighting their limitations in meeting the essential criteria for open sharing of IoT MUD data within the Internet community.

\textbf{Data Marketplace for IoT Devices:} 
Prior research proposed decentralized marketplaces of IoT data using blockchain and public ledger technologies \cite{GSRamachandranISC22018}. Their primary objective has been centered around  sharing the data measured or generated by IoT devices, while focusing on aspects like data privacy \cite{PKlaineDLT2023, HSubramanianJMIR2023, VKoutsosTDSC2022, LGiarettaICDEW2021,FYangTII2022}, data sovereignty \cite{VSSADaliparthiEICC2023, PGuptaJNCA2022}, security of private queries \cite{MZhangFGCS2023}, scalability \cite{GVictorIoT2023}, and data pricing \cite{SAAzcoitiaDE2022, MZhangTBD2023}. Researchers also considered scenarios when IoT-generated data is shared through a specific type of network, such as software-defined wireless body network \cite{KHasanCN2022}, a network connecting autonomous governmental authorities \cite{YLiuICBCT2021}, a network connecting healthcare systems \cite{JLeeJISA2022, RKumarTII2022, JZhangTII2022}, cloud-based IoT networks \cite{TLiIoTJ2022, YZhangJSAC2022, SDOkegbileIoTJ2022, ZZhouTII2023}, transportation systems network \cite{SJiangIS2023}, vehicle networks \cite{JHuangTSN2023}, and metaverse networks \cite{HThienFGCS2023}. 
Our work differentiates itself by its objective of sharing data of expected behaviors (\ie MUD files) of IoT devices. Therefore, our architecture is driven by specific design choices such as market openness, data transparency, data quality measures, and cost-effectiveness for large MUD data.

\textbf{MUD Data for IoT Network Security:}
MUD, a structured data format containing permitted IoT network communications, has attracted significant interest from industry and academia since its formal inception in 2019 \cite{rfcMUD}. 
MUDScope \cite{LMZangrandiACSAC2022} detects persistent (lasting long) unexpected IoT communications that do not conform to MUD profiles.
MUDIS \cite{ABremler-BarrNOMS2022} highlights variations of a MUD profile caused by deployment settings and proposes a method to generalize MUD profiles to cater to a variety of deployments.
PicP-MUD \cite{APashamokhtariWoWMoM2022} proposes to extend the standard MUD specifications by incorporating characteristics of traffic payloads communicated by individual MUD flows, enabling fine-grained security monitoring.
Works in \cite{SAHarishNOMS2023} developed methods that use programmable data-plane switches to automatically enforce MUD-specified policies for IoT devices connected to edge networks.
Though MUD is effective and enables various network security operations, its data supply is critical. This paper aims to accelerate the adoption of MUD within the cybersecurity community by developing a special decentralized marketplace that enables the open sharing of MUD files for IoT devices that become available in the market and are deployed in various networks.

\textbf{Commercial Centralized Data Marketplace:}
One may contemplate utilizing centralized data marketplaces available today for potential sharing and/or trading IoT MUD data within the network security community. These data marketplaces are either operated by major technology providers like Amazon AWS Data Exchange \cite{AWSDataExchange}, Google Cloud Marketplace \cite{GoogleMarketPlace}, Red Hat Marketplace \cite{RedHatMarketPlace}, or innovative startups like Databricks Marketplace \cite{DatabricksMarketplace}.
Despite their diverse business models and data-sharing mechanisms, the centralized data marketplaces currently available have limitations. They often fall short in terms of openness, transparency, and cost-effectiveness for sharing and reviewing MUD data in the security community. These existing marketplaces are primarily driven by profit motives, contrasting with the non-profit, open-source nature of sharing IoT MUD profiles within the security community.
To address these shortcomings, we design a dedicated data marketplace tailored specifically for IoT MUD data. Our proposed marketplace is accessible to the entire community, allowing various security interest groups and individual enthusiasts to contribute immutable MUD data for different IoT devices. Importantly, it upholds the open-source ethos of the data, and costs are covered by the first requester with the most urgent demand. This approach ensures sustainability and promotes a collaborative effort to systematically enhance Internet security.

\begin{figure*}[t!]
	\centering
	\includegraphics[width=0.7\linewidth]{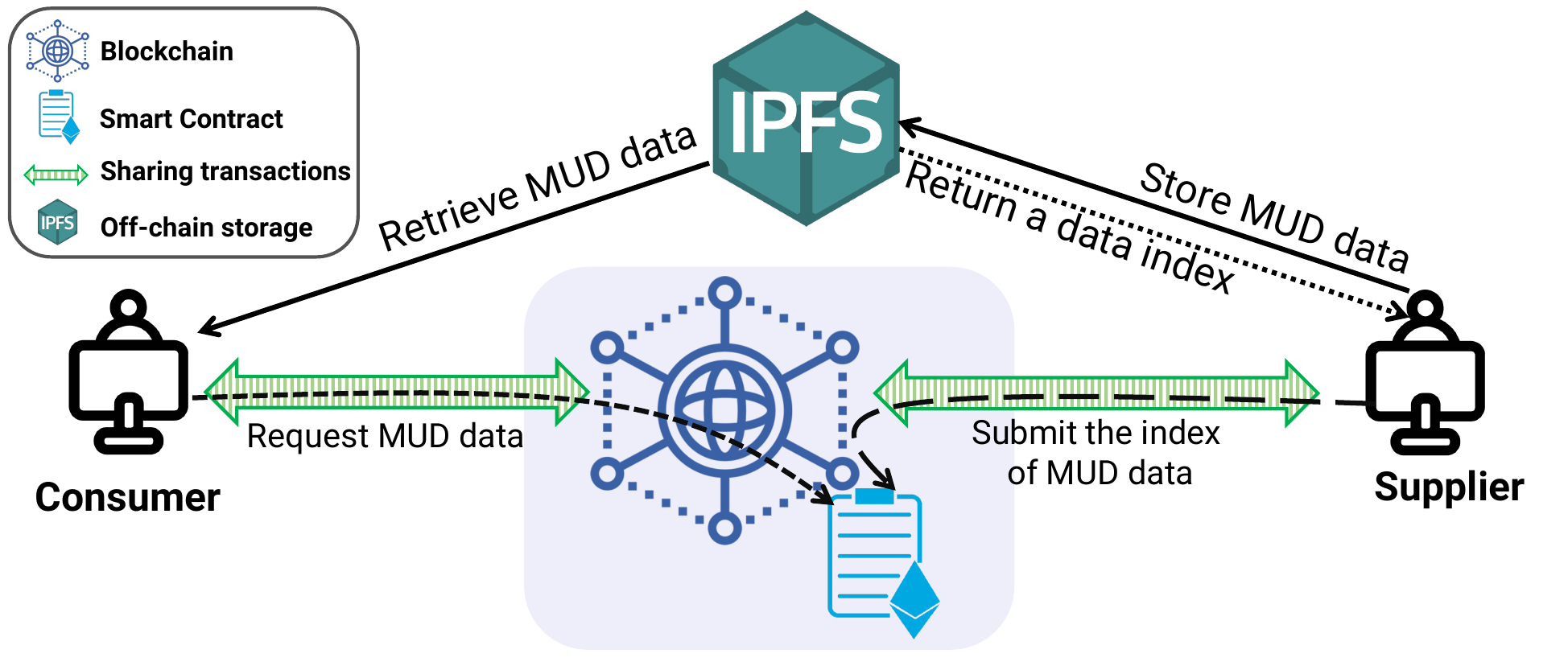}
	\caption{System architecture of the data marketplace for exchanging IoT MUD data.}\label{fig:architecture}
    \vspace{-3mm}
\end{figure*}

\section{Decentralized Marketplace of \\IoT Behaviors Data}\label{sec:tool}

In this section, we begin by highlighting the importance and requirements of sharing data of IoT expected behaviors within the cybersecurity community (\S\ref{sec:DesignRationales}). We next discuss our design of a decentralized data marketplace (\S\ref{sec:systemArchitectureProcess}).

\subsection{Sharing  Data: Why and How?}\label{sec:DesignRationales}
The maturity of IoT manufacturers and/or cloud providers can vary widely. This makes the IoT market particularly vulnerable to low-quality manufacturing and high-risk communication practices. Standards like the IETF RFC8520 encourage manufacturers to specify the expected usage of their devices on the network (\ie baseline network behaviors) in machine-readable data models (aka, MUD files) to effectively protect IoT devices at network levels and reduce their attack surfaces. However, manufacturers are not mandated or at least sufficiently incentivized \cite{rfc9238} to publish necessary MUD data records.

Note that MUD files do not include private or sensitive information; the original standard anticipated manufacturers to make this data public.
MUD files may become publicly available by large and mature manufacturers for some of their premium IoT products, or by researchers who experiment with representative IoT devices in their labs \cite{VAndalibiACSAC2021}. However, most IoT devices will likely remain unmanaged, lacking data on their expected behaviors (\ie MUD files). Even those with a MUD file, their data can still be incomplete or not up-to-date \cite{NISTIR8349}. 
The expected behaviors of a  given IoT device type may vary slightly depending on its deployment network configurations or firmware versions \cite{ABremler-BarrNOMS2022}.
In addition, there is little incentive for professionals in the IoT security community to proactively update and share the MUD files they may be able to generate.

\textbf{Open, Transparent, and Decentralized Data Sharing:}
Currently, IoT MUD data (expected behaviors), at best, can be made available via isolated scattered repositories \cite{MUDcheck2022} in an ad-hoc manner, thereby limiting the benefits for both consumers and suppliers of data.
Therefore, more than ever, a data marketplace that encourages the cybersecurity community to share IoT behavior data (potentially for some monetary rewards) is needed. It is important to note this marketplace cannot emerge and sustain itself by a single authority. Given the heterogeneity of devices in conjunction with the dynamics in the expected behaviors, multiple and independent data suppliers (\eg manufacturers, security vendors, system integrators, network operators, certification labs, or researchers) are unavoidable. This will entail permissionless, instead of permitted, suppliers. On the other hand, every consumer expects to discover and/or request the MUD file of any IoT devices (\eg make/model), possibly with certain firmware versions.
These operations demand two critical properties of a desired marketplace: market openness and data transparency. Public users and stakeholders (instead of a confidential group) will need the ability to obtain full knowledge about suppliers, consumers, and transactions while allowing them to perform independent audits. Therefore, the exchanged MUD data should be available in an unencrypted format. To realize full transparency, particularly at scale, the marketplace must be decentralized \cite{GVictorIoT2023}.
Moreover, the open-source nature of the shared IoT MUD files ensures that the data is accessible to the community after its costs have been covered by the initial requester, who urgently requires the updated MUD profile. This stands in contrast to proprietary asset holding. Centralized marketplaces typically do not support the open-source feature, while public repositories lack mechanisms to incentivize data providers by compensating the costs associated with MUD data generation.

\textbf{Measuring Data Quality:}
Unlike non-fungible tokens (NFT), text-based MUD files are not unique and can be generated by any entity. Therefore, estimating the quality (in the absence of authenticity) of MUD data exchanged on the marketplace, or at least quantifying supplier reliability, by consumers is crucial.
This aspect becomes even more critical in a permissionless data market where data consumers can discover/request MUD files supplied by community contributors, not necessarily by known or trusted (certified) suppliers.
One may argue that a larger MUD file implies more (fine-grained) allow-listed rules and, thus, better quality. 
However, having more allowed rules, if unnecessary (\eg outdated ones due to firmware upgrades), can lead to an expanded attack surface. On the other hand, having fewer rules may undesirably limit certain functionalities of the intended device.
Therefore, we need an indication of data quality (\eg measure of the supplier reputation) to maintain a healthy ecosystem of MUD data sharing.

\textbf{Costs of Sharing Large MUD Data:} MUD files are JSON-formatted text files, and their size can vary between several KBs to tens of KBs, much larger than data sizes that can be cost-effectively exchanged on decentralized platforms \cite{AGervaisCCS2016}. Therefore, options like off-chain storage \cite{JJayabalanJPDC2022} or compression \cite{SSArslanCC2022} may be required for our decentralized data marketplace without compromising other requirements.

\subsection{System Architecture and Process}\label{sec:systemArchitectureProcess}

Based on the above-mentioned requirements, we now design a decentralized data marketplace. We use blockchain to record automatic transactions enabled by a specialized smart contract between consumers and suppliers. Also, We employ off-chain storage to host relatively large MUD data, managing transaction costs on blockchain.
We discuss the marketplace architecture in \S\ref{sec:arch}, transaction events (functions) in the smart contract in \S\ref{sec:smartContract}, and deadline settings to manage the timeline of events in \S\ref{sec:EventTimeline}.

\subsubsection{Blockchain-based System Architecture}\label{sec:arch}
Our data marketplace is designed to allow consumers (\eg administrators of IoT networks) to request and purchase MUD data of specific devices from any potential suppliers (\eg device manufacturers,  researchers, or managed security providers).
Blockchain is an appropriate choice of technology to address our problem needs. It guarantees the immutability of the exchanged data and provides decentralized management of the marketplace. 

Broadly speaking, blockchain can be realized in two ways: permissioned and permissionless. The permissioned ones are strictly accessed by certain authorized users, while the permissionless ones are publicly accessible. Therefore, we use the Ethereum blockchain in our design as it is permissionless and supports custom smart contracts for data transactions. 
{Noting that Ethereum is a programmable blockchain that supports third-party developers in deploying customized decentralized applications, it has been widely adopted in prior works.} 
Fig.~\ref{fig:architecture} illustrates our system architecture with Ethereum as the foundational module. This module stores necessary programmable interfaces (\ie smart contract) to support data exchanges between consumers and suppliers, which will be discussed in \S\ref{sec:smartContract}. 

As discussed in \S\ref{sec:DesignRationales}, qualified MUD data files are often sized about tens of kilo bytes, which can be fairly expensive (not cost-effective) \cite{ALaurentBRA2022} to exchange on the Ethereum blockchain. To guarantee the immutability of MUD data while managing the cost of transactions, we employ an off-chain storage method called InterPlanetary File System (IPFS) \cite{DTrautweinSIGCOMM2022}, which returns a small index (less than 1KB) of the stored MUD file by hashing its entire content. 
Notably, the open-source community purposefully crafted IPFS to offer a secure and distributed storage service for peer-to-peer networks. This makes IPFS the fitting off-chain storage solution for our data marketplace.
Manipulating a stored MUD file on IPFS will result in a different hash index, thus, ensuring data immutability.
As shown in Fig.~\ref{fig:architecture}, the actual MUD data will be uploaded by the supplier (following a consumer request) and can be retrieved by the consumer from IPFS using the hash index of MUD data will be exchanged in a transaction on the blockchain. The step-wise process of sharing MUD data is discussed next.

\subsubsection{Smart Contract for Exchanging MUD Data}\label{sec:smartContract}

\begin{figure}[t!]
	\centering
	\includegraphics[width=\linewidth]{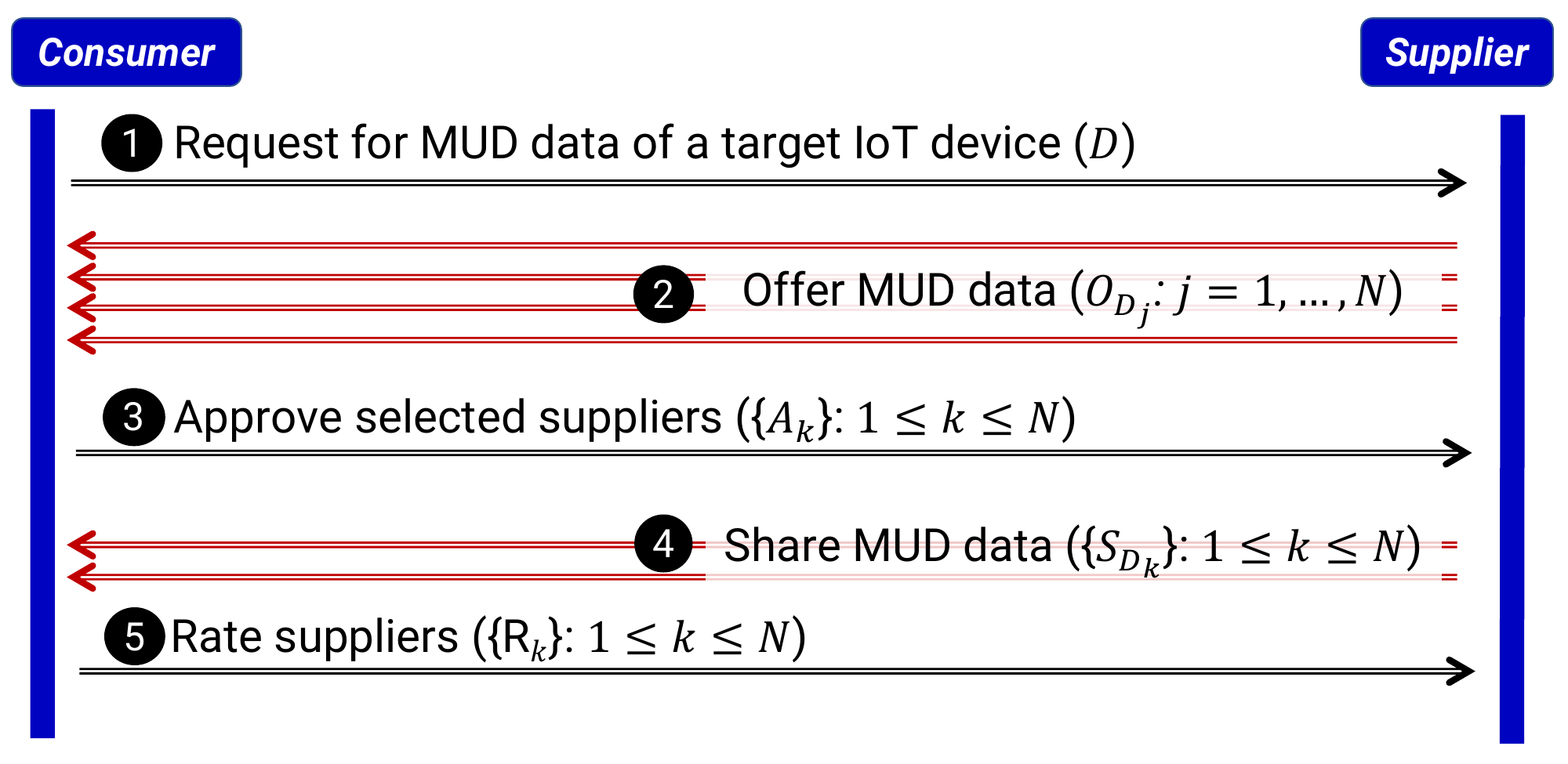}
	\caption{Sequence of transactions (events) between consumer and supplier for exchanging data of IoT expected behaviors (MUD file).}
	\label{fig:flow}
\end{figure}

A smart contract is a specialized program on a blockchain that facilitates interactions between consumers and suppliers \cite{PTolmachCS2022}, each uniquely identified by an (Ethereum) address. In the context of our marketplace, as detailed in \S\ref{sec:arch}, the smart contract has been carefully crafted to enable the sharing of the MUD data index rather than the raw data. This strategic choice is primarily aimed at optimizing cost-effectiveness while ensuring the preservation of immutability. The transactional process, illustrated in Fig.~\ref{fig:flow}, unfolds through a series of five distinct steps: request, offer, approval, data sharing, and rating. In what follows, we delve into each of these steps individually.

\textbf{Request:}
An entire MUD data exchange transaction begins with a request published by the consumer. This request is made openly available on the blockchain, allowing potential suppliers to review its contents.
The consumer is expected to specify the intended IoT device within a request, denoted by $D$ in step \ciao{1} of Fig.~\ref{fig:flow}. These specifications contain details such as the manufacturer, model, hardware variant, and firmware version. Prospective suppliers utilize these device specifications to determine whether they can offer the expected MUD data. This selective approach ensures the preservation of data quality within the marketplace and contributes to maintaining the suppliers' reputations.
In addition to device specifications, the consumer's request includes a designated budget. This is a reference for potential suppliers, aiding them in tailoring their offers to align with the consumer's financial constraints.
To facilitate tracking and accessibility, each request posted on the marketplace is automatically associated (by blockchain) with a unique identifier (UID). This UID is generated by hashing the current block timestamp combined with the consumer's address, thereby guaranteeing its uniqueness. The availability of such UIDs enables stakeholders---including consumers, suppliers, and interested third parties---to locate the corresponding transaction on the decentralized ledger easily. Moreover, these stakeholders can access the historical MUD data exchanged, albeit possibly outdated, and make informed assessments regarding the reliability of specific suppliers.

\textbf{Offer:}
Upon publication of a request on the marketplace, all prospective MUD data suppliers (referred to as $j$ in step \ciao{2} of Fig.~\ref{fig:flow}) have the opportunity to present their respective offers ($O_{D_{j}}$) for the expected IoT behavioral profile. These offers contain metadata, descriptions, and pricing details.
An offer's metadata and description components encompass quantifiable data points. These may include factors such as the count of ACE\footnote{Access Control Entry} flows, the MUD file size (KB), the scope of incorporated flows, and the network setting in which the supplier generated the MUD data.
Each supplier can only make a single offer to a request, upholding data quality standards. This ensures that suppliers submit what they consider to be the most appropriate match.
Moreover, each offer is accompanied by its price (in Ether\footnote{Ether is the native cryptocurrency for the Ethereum blockchain and network of miners.}). While the offered price does not have to be lower than the consumer budget, if it is, the supplier could enhance their chances of being chosen. This pricing mechanism reflects the currency's circulation and utilization within the broader Ethereum ecosystem.

\textbf{Approve:}
Within the sharing transaction process, a consumer is empowered to review, select, and subsequently approve none, one, or multiple suppliers (referred to as $A_k$ in step \ciao{3}) from $N$ data offers. 
It is worth noting that the intrinsic value of a MUD profile is tied to the textual data content encoded in JSON format. To safeguard the interests of suppliers, the (selection and) approval phase is done without directly inspecting (viewing) the actual content of intended MUD data. 
Consumers, instead, should estimate the data value based on each offer's metadata and descriptive information.
To address uncertainty or trust concerns, suppliers' reputations or ratings (discussed in the forthcoming function) can be assessed by consumers based on historical transaction records on the blockchain.
Therefore, the selection can be made by (optimally or deterministically) incorporating factors such as pricing/budget considerations, metadata, and/or supplier reputation. Combining safeguards and open records enhances the consumer's ability to make an informed selection, thereby balancing the interests of consumers and suppliers in the marketplace.

\textbf{Share:}
Upon approving (selected) supplier(s), the expected payment (according to the offered prices) is made from the consumer to the corresponding suppliers. Following that, the selected supplier(s) will share the IPFS hash index of the intended MUD data ($S_{D_{k}}$) on the blockchain.
Practically, suppliers may not be able to share the MUD data immediately---some delays are unavoidable. To manage expectations, we will introduce the notion of explicit ``deadline'' later in \S\ref{sec:EventTimeline}.
The hash index enables the consumer to retrieve the corresponding MUD data from the IPFS storage system.

\textbf{Rate:}
Finally, after the shared MUD data is retrieved from the IPFS database, the consumer will have the opportunity to give a rating to individual suppliers (step \ciao{5}) based on the quality of their data record(s). Numerical rates (\eg percentage or binary value) to suppliers are made publicly available on the blockchain. They can be used to measure the trust/reputation of suppliers in future transactions. 
With the rating mechanism in place and appropriately utilized by many consumers over time, prospective marketplace users become empowered with a metric to assess the quality of a certain supplier and their shared data records.
We note that malicious users may attempt to arrange ``scam'' reviews for themselves that came from illegitimate accounts, misleading the marketplace users. We will see later in \S\ref{sec:gasConsumption} that such malicious activities can be prohibitively expensive at scale given that each transaction's cost (gas consumption) can be in the order \$100. 
Also, the shared MUD data will be publicly available on the blockchain to every node to review. Therefore, detecting blatantly inappropriate data records will not be impossible. 

\subsubsection{Time Management in Sequence of Events}\label{sec:EventTimeline}
So far, we have discussed individual functions (events) our smart contract enables. Note that these events unfold in a sequential manner. However, some events may never occur or occur with unknown delays (reasonable or not). These uncertainties could be anticipated by consumers and suppliers. For example, a request might receive no responses from suppliers, or a supplier might face challenges in sharing the approved MUD data. 
Therefore, consumers and suppliers would find it beneficial to enforce deadlines for their requests, offers, and approvals.
This proactive approach protects their interests, considering factors such as responsiveness or financial constraints. By setting deadlines, parties involved can better navigate uncertainties and ensure a more predictable and controlled process within the decentralized setting.

Therefore, each function, except for ``rate'', incorporates a deadline value specified by the caller user. The deadline for ``rate'' is a built-in value defined in the smart contract itself.
A consumer patiently awaits the deadline to select offers from potential suppliers. Likewise, every offer is accompanied by a supplier-specified deadline within which the consumer must make a decision.
Upon approving an offer via the ``approve'' function, if no submission occurs as intended by the ``share'' function, the Ether is reimbursed to the consumer. Furthermore, the deadline specified in the ``submit'' function signifies when the consumer should either rate the transaction or allow the default value to persist.
This strategic use of deadlines effectively prevents unfinished steps within a transaction that could compromise the interests of the parties involved. By adhering to these time limits, the system ensures smoother interactions.

\begin{figure*}[t!]
	\centering
	\includegraphics[width=0.67\linewidth]{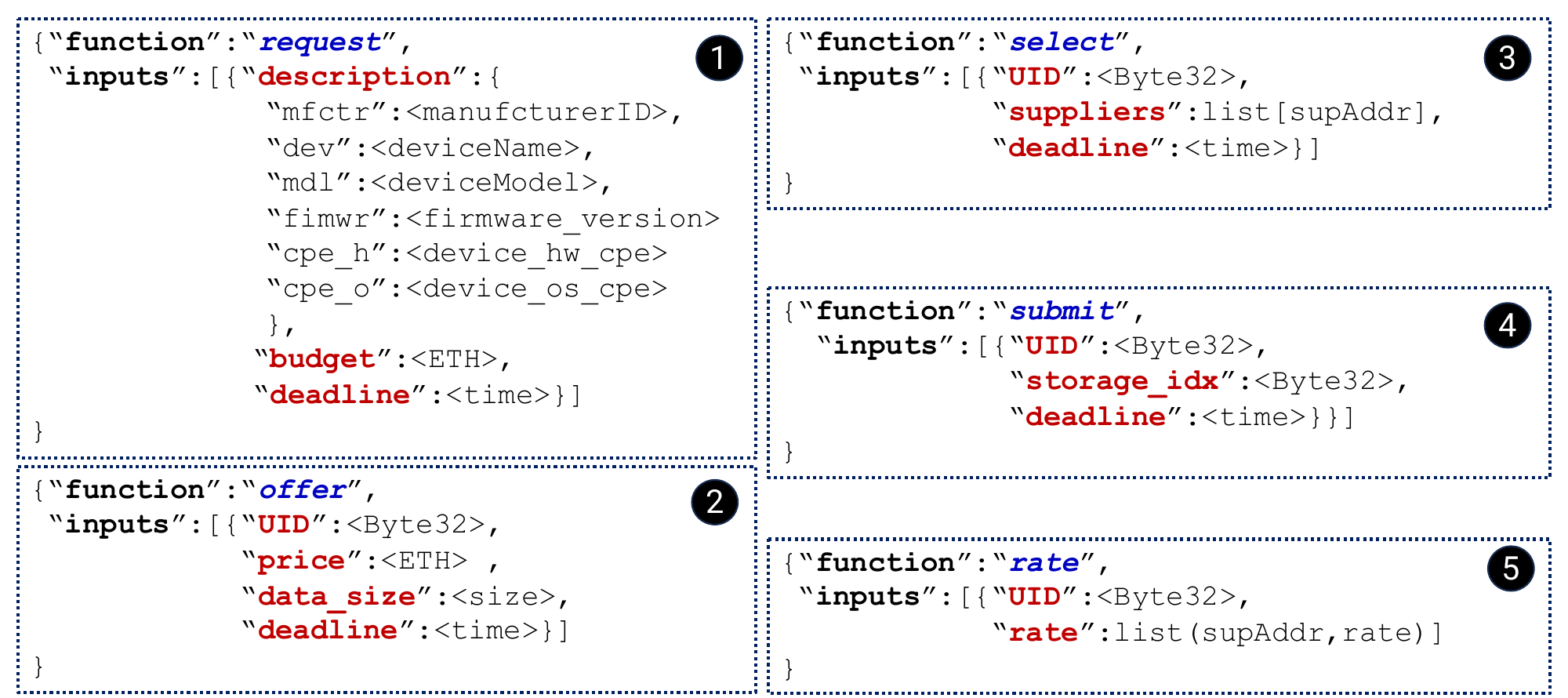}
	\caption{Five concrete functions facilitate blockchain data exchange between consumers and suppliers.}	\label{fig:API}
    \vspace{-3mm}
\end{figure*}

\section{Implementation and Evaluation}\label{sec:eval}

This section starts by outlining our prototype implementation on a private Ethereum blockchain and detailing the experimental setup  (\S\ref{sec:prototype}).
Subsequently, we delve into the discussion of eleven distinct testing scenarios. These encompass basic transactions, transactions characterized by varying selection strategies a consumer may choose, cases involving exceptions, and the exploration of past transactions. These scenarios collectively serve to illustrate the seamless functionality of our data marketplace prototype (\S\ref{sec:demonstrativeScenarios}).
We also benchmark the cost of executing each function of the MUD data-sharing smart contract to demonstrate the financial viability of our proposed marketplace (\S\ref{sec:gasConsumption}).

\subsection{Prototype Environment and Experimental Setup}\label{sec:prototype}
We realized our implementation of the decentralized data marketplace, designed for exchanging IoT MUD data, using Ganache version 7.9.0 \cite{Ganache}. 
{Ganache is a widely used software tool in the Ethereum and blockchain development ecosystem. It is a private emulator, often referred to as a ``blockchain sandbox'' that allows developers to create and test Ethereum applications in a controlled and local environment. }
We have utilized the default configuration of 10 users, each having 100 Ethers assigned. As noted by prior work \cite{SSingh2022}, a blockchain-based system validated effectively within the Ganache environment can be seamlessly transitioned to the public blockchain without necessitating additional engineering modifications.
We developed the data-sharing smart contract using Solidity version 0.8.18 \cite{Solidity}, an object-oriented, high-level language for implementing smart contracts within the Ethereum blockchain.
Regarding the off-chain IPFS storage of MUD data, we established a local instance \cite{IPFS} on the same machine hosting the private Ethereum blockchain, facilitating an integrated setup.
We employ the standard Web3 RPC/JSON interface \cite{Web3RPC} for interacting with different nodes within our private blockchain. This facilitates the execution of smart contract functions and/or access to the ledger.

Fig.~\ref{fig:API} presents a detailed representation of the specific syntax for the five functions we implemented for our MUD data-sharing smart contract, along with their corresponding parameters.
As discussed earlier in \S\ref{sec:smartContract}, a consumer user initiates the process by invoking the {\myverb{request}} function, providing certain parameters to describe the device\footnote{While providing more information helps narrowly specify the expected data, some fields may be left blank (consumer options).} for which the MUD data is requested, along with indicating the budget and expected deadline.   
In addition to the manufacturer name, device name, device-specific model number, and firmware version, the consumer may choose to provide a structured naming scheme like CPE (Common Platform Enumeration)\footnote{The code ``{\myverb{cpe:2.3:o:amazon:echo\_firmware:2018-04-27:* :*:*:*:*:*:*}}'' refers to Amazon Echo OS/Firmware 2018-04-27.}, currently maintained by the National Institute of Standards and Technology (NIST) \cite{CPE2023} as part of its U.S. National Vulnerability Database (NVD). Similarly, the other four functions (\ie {\myverb{offer}}, {\myverb{select}}, {\myverb{submit}}, and {\myverb{rate}}) are implemented according to our design requirements (\S\ref{sec:smartContract}).

\begin{table*}[!t]
	\caption{Four groups of illustrative test scenarios on MUD data of three representative IoT devices: Amazon Echo, LIFX lightbulb, and Samsung smart camera.}
	\label{tab:demonstrativeScenario}

	\begin{adjustbox}{width=\textwidth}

        \renewcommand{\arraystretch}{1.1}
        \begin{tabular}{|l|l|l|l|l|l|l|}
        \hline
         \rowcolor[rgb]{ .906,  .902,  .902}   \textbf{Scenario}  & \textbf{Consumer} & \textbf{Suppliers} & \textbf{\# selected} & \textbf{\# submitted} & \textbf{\# rated} & \textbf{Specification}            \\ \hline\hline
        BS01           & U1          & \{U7\}          & 1                & 1                 & 1             & Full cycle of five functions.                \\ \hline
        BS02           & U2         & \{U7\}          & 1                & 1                 & 0             & BS01, without the rating transaction.           \\ \hline
        BS03           & U3          & \{U7, U8, U9, U10\}           & 1                & 1                 & 1             & BS01 with four offers; the first arrival is selected.              \\ \hline
        BS04           & U4          & \{U7, U8, U9, U10\}          & 3                & 3                 & 3             & BS03; the first three arrivals are selected.             \\ \hline\hline
        ES01           & U5          & \{U7, U8, U9, U10\}         & 3                & 0                 & 0             & No submission is made.           \\ \hline
        ES02           & U1          & \{U7, U8, U9, U10\}        & 0                & 0                 & 0             & No offer is selected.            \\ \hline
        ES03           & U2          & \{\}          & 0                & 0                 & 0             & No offer is supplied.            \\ \hline\hline
        SS01           & U3          & \{U7, U8, U9, U10\}          & 3                & 3                 & 3             & The top three largest sizes (within budget) are selected.          \\ \hline
        SS02           & U4          &   \{U7, U8, U9, U10\}        & 3                & 3                 & 3             & The lowest three prices (within budget) are selected.          \\ \hline
        SS03           & U5          & \{U7, U8, U9, U10\}         & 3                & 3                 & 3             & Highly reputable suppliers are selected.            \\ \hline\hline
        VS01          &  U6          & ---          & ---                 & ---                 & ---              & User viewing previous transactions. \\ \hline
        \end{tabular}
        \end{adjustbox}
\end{table*}

\subsection{Illustrative Scenarios}\label{sec:demonstrativeScenarios}

Within our controlled experimental environment, we have developed Python3 scripts\footnote{A GitHub repository containing our source codes and MUD files is publicly available at \cite{shareMUDrepo2023}.} that leverages the Web3 library. These scripts automate the testing of illustrative scenarios, thereby effectively showcasing the operational aspects of our prototype.
In our experiments, consumers are chosen from users U1 to U6, while suppliers are drawn from users U7 to U10.
The scripts execute eleven scenarios focused on three representative IoT MUD files: Amazon Echo, LIFX lightbulb, and Samsung smart camera. These MUD files have been sourced from a public dataset \cite{AHamzaIoTSP2018}.
For each MUD file, we created four versions to encapsulate varying qualities: (i) high quality (HQ) corresponds to a MUD file that contains the complete list of the original ACEs \cite{AHamzaIoTSP2018}, (ii) medium quality remote (MQR) corresponds to a MUD file that contains the complete list of the original ACEs, excluding all local rules, (iii)  medium quality outbound (MQO) corresponds to a MUD file that contains the complete list of the original ACEs, excluding all inbound rules, (iv) low quality (LQ) corresponds to a MUD file that contains the complete list of the original ACEs, except for a ``randomly-selected'' rule.
During experimental transactions, the participating supplier randomly chooses one of these four versions in each scenario. 
Also, the gas consumption for each function call is recorded during all test instances.

Table~\ref{tab:demonstrativeScenario} summarizes our eleven scenarios under four groups: basic scenarios (BS01-BS04), exception scenarios (ES0-ES3), selection scenarios (SS0-SS03), and the viewing scenario (VS01). Each scenario is designed to come with random consumers and suppliers, as well as the number of selected, submitted, and rated suppliers. 

\begin{figure*}[t!]
	\centering
	\includegraphics[width=0.7\linewidth]{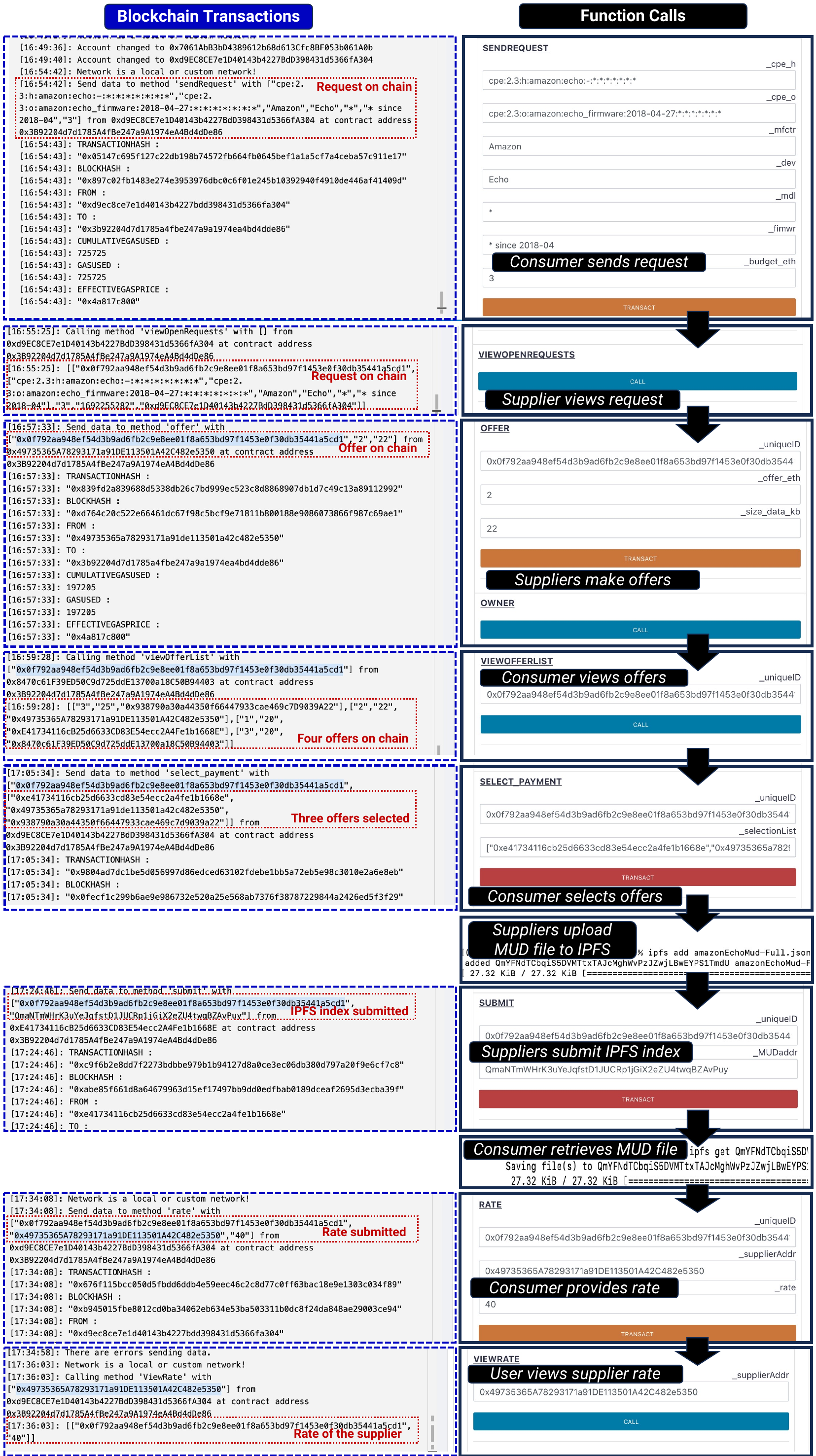}
	\caption{Transactions made in the scenario BS04.}
	\label{fig:Scenario3}
    \vspace{-3mm}
\end{figure*}

\textbf{Basic scenarios:}
In this group, we aim to demonstrate baseline transactions between data consumers and suppliers on our private blockchain completed before respective deadlines. Let us start with BS01 to showcase a full cycle of five functions, where consumer U1 initiates a request. Supplier U7 promptly responds and is subsequently chosen by U1. Following this, U7 submits the requested MUD file data (of HQ quality) and receives a favorable rate (\ie 100\% in our implementations) from U1.   
In the second basic scenario, we replicate the steps of BS01, except that the consumer U2 opts not to provide a rating for supplier U7.
BS03 differs from BS01 by showcasing a scenario whereby four offers are made but U3 selects the first arrival. The last basic scenario, BS04, builds upon BS03 by extending the selection to the first three arrivals---submissions are of HQ, MQO, and LQ quality that are randomly selected by our testing program and rated accordingly (100\% for HQ, 40\% for MQO, and 10\% for LQ). 
Fig.~\ref{fig:Scenario3} visualizes the steps involved in the basic scenario BS04 about the Amazon Echo data---screenshots are from our prototype environment. 
The right column, shown by solid black frames, lists the sequence of functions the consumer and suppliers invoked via the Web3 interface. The left column, shown by dashed blue frames, lists the corresponding transactions on the blockchain---we highlighted key messages with dotted red sub-boxes.
These screenshots encompass all possible operations that users might execute. This includes a consumer initiating requests with expected parameters, suppliers viewing open requests, suppliers submitting offers, consumers examining and selecting offers, chosen suppliers uploading MUD data to the IPFS database, suppliers submitting IPFS indexes, consumers retrieving MUD files based on corresponding IPFS indexes, consumers providing ratings, and other users accessing completed transactions and the ratings of specific providers.

\textbf{Exceptions scenarios:}
We developed three scenarios (on top of BS04) to test how our smart contract handles exceptions (\eg deadlines expire). These scenarios entail disrupting the chain of transactions midway---an intermediate step gets omitted. In ES01, three of the four offers are selected by U5, yet none of the corresponding suppliers proceed with their submissions. In this situation, our smart contract takes action by concluding the incomplete transaction and refunding the Ether amount initially committed by the consumer node. ES02 illustrates a scenario in which the requesting consumer selects no offer. This could be due to factors such as high prices, small data sizes, or poor supplier ratings. ES03 highlights the case where no offer is made due to factors like a lack of data or the consumer indicating small budgets for the requested data. 

\begin{table*}[!t]
    \centering
	\caption{Average size and cost of smart-contract functions for sharing MUD data.}
	\label{tab:GasConsumption}
	\begin{adjustbox}{width=0.8\textwidth}
        \renewcommand{\arraystretch}{1.1}
        \begin{tabular}{|l|l|l|l|l|}
        \hline
         \rowcolor[rgb]{ .906,  .902,  .902} \textbf{Function} & \textbf{Data size on blockchain} (Byte) & \textbf{Gas consumption} & \textbf{Transaction fee} (ETH) & \textbf{Cost} (USD) \\ \hline
        Request  & 709                                      & 736,408         & 0.017165673            & 28.02                \\ \hline
        Offer    & 262 & 199,358         & 0.004647035            & 7.58                \\ \hline
        Select   & 388                                     & 173,681          & 0.004048504            & 6.60                 \\ \hline
        Submit   & 301                                           & 218,216         & 0.005086615            & 8.30                 \\ \hline
        Rate     & 143                                            & 140,474          & 0.003274449            & 5.34                 \\ \hline
        \end{tabular}
    \end{adjustbox}
\vspace{-2mm}
\end{table*}

\textbf{Selection scenarios:}
Three scenarios are developed to showcase a variety of strategies (beyond just considering arrival times) that can be employed by consumers when interacting with suppliers. 
In these scenarios, the respective consumer selects based on data size (\ie the larger is assumed to be higher quality) in SS01, offered prices (\ie the lowest price is more economical) in SS02, and supplier ratings (\ie the more reputable, the more trustable) in SS03.

\textbf{Viewing scenario:}
Lastly, in VS01, no transaction occurs between users. Instead, we demonstrate how any user within the data marketplace is able to openly access previous transactions. For this specific scenario, U6 views records on the blockchain, including those MUD files (IPFS indexes) previously shared in the past. This highlights the transparency and openness of our proposed marketplace.

\subsection{Quantifying Transaction Costs}\label{sec:gasConsumption}
After verifying the functionality of our marketplace prototype, we focus on financial costs.  
Every operation conducted on the Ethereum blockchain, facilitated by the smart contract functions, will involve financial costs (\ie gas consumption\footnote{Gas is a unit of measure for computational work and resources required to execute and validate transactions on the Ethereum network \cite{1AAZarirTSEM2021}.})). These costs (paid in the form of Ether) are required by distributed miner nodes for the validation and inclusion of the transaction in the public ledger.

During our illustrative scenarios, we recorded the data size on the blockchain and the associated gas consumption of each function call. Subsequently, we computed the respective transaction fee (in Ehter) and its equivalent value in real-world currency (USD).
At the time of our experiments (23 August 2023), $4.3e7$ gas = 1 Ether = 1633 USD. 

Table~\ref{tab:GasConsumption} summarizes the average on-chain data size and gas consumption per each function of our smart contract, as measured during illustrative scenarios. As an example, considering the request function (first row), a total of 709 bytes are transmitted to the Ethereum ledger. This action consumes 736,408 gas, equating to 0.017 Ether, which is valued at approximately 28.02 USD.

It is evident that executing a complete cycle of five functions for the exchange of a MUD file would result in a total cost of less than 100 USD. This expense is divided between the consumer (request, select, and rate functions) and the supplier (offer and submit functions). It is important to note that this amount is directed towards compensating the network of miners and does not factor in the price of data that individual suppliers might explicitly request in their offers.
That said, the expense associated with marketplace transactions to acquire a MUD file remains reasonable, especially for larger enterprises managing several IoT device types but each in significant quantities (\eg hundreds or thousands).   
Given the investment made to procure and construct IoT networks (with enterprise IoT devices such as cameras, lightbulbs, and air quality sensors costing around a few hundred dollars each), allocating a portion of that investment towards utilizing this marketplace for obtaining high-quality data can help fortify the security of these vulnerable infrastructures.

\section{Discussion}\label{sec:discuss}
To our knowledge, this paper is the first attempt to conceptualize a decentralized data marketplace dedicated to sharing IoT MUD data. While our efforts lay the foundation, certain aspects deserve additional refinement, which is beyond the scope of this paper and thus left for future work.

Firstly, as the marketplace gains momentum, a request for data offers is likely to attract many choices, each with unique features like suppliers, prices, deadlines, and data quality. 
Selecting from these options could be challenging for the caller consumer. To help consumers choose the best subset based on their needs, using optimization algorithms \cite{21IoTJwitnessBC} can be beneficial.
Secondly, our rating function employs a 100-point metric supplied by the respective consumer. While this approach is relatively common in digital marketplaces \cite{PChen2013}, a more detailed rating system with additional descriptive information could help prospective consumers make better decisions when selecting offers from suppliers with past records.
Thirdly, deadlines in our current design address scenarios where legitimate users opt not to complete the entire transaction (\eg a consumer choosing none of the available offers or an approved supplier failing to share the expected data by the set deadline). We acknowledge that an open data marketplace is likely to draw malicious actors with diverse motivations to disrupt and undermine the ecosystem through actions like fraudulent requests, spam ratings, and repetitive offers without actual sharing. Supplementary security measures against potential attacks are also expected to be developed as future works to ensure the marketplace's resilient functioning on a public network like Ethereum.

\section{Conclusion}\label{sec:conclude}
Despite the proven effectiveness of Manufacturer Usage Description (MUD) data in enhancing IoT network security by allowlisting expected device behaviors, its widespread adoption encounters hurdles due to the cumbersome process of sharing high-quality MUD data within the cybersecurity community.
This paper introduced and implemented an open and decentralized marketplace for the exchange of MUD data using blockchain technology and smart contracts.
We outlined the prerequisites for a data marketplace intended for the exchange of IoT MUD files and subsequently developed a decentralized architecture utilizing the Ethereum blockchain. This architecture is realized through a sequence of five specific functions (request, offer, select, submit, rate), each executed with specific parameters.
Next, we translated our design into a functional prototype within a private chain environment. We proceeded to showcase diverse interactions between distributed consumers and suppliers through a series of illustrative scenarios. Our experimental results underscore the cost-effectiveness of the proposed marketplace.

\balance

\bibliographystyle{IEEEtran}
\bibliography{ProtoIoT}

\end{document}